\documentclass{osa-article}

\journal{osajournal}

\articletype{Research Article}

\usepackage[export]{adjustbox}
\usepackage{amsmath}
\usepackage{amssymb}
\usepackage{bm}

\begin{document}

\title{Microwave-to-optical conversion using lithium niobate thin-film acoustic resonators}

\author{
Linbo Shao,\authormark{1,6} 
Mengjie Yu,\authormark{1}
Smarak Maity,\authormark{1}
Neil Sinclair,\authormark{1, 2}
Lu Zheng,\authormark{3}
Cleaven Chia,\authormark{1}
Amirhassan Shams-Ansari,\authormark{1}
Cheng Wang,\authormark{1,4}
Mian Zhang,\authormark{1,5}
Keji Lai,\authormark{3} 
and Marko Lon\v{c}ar\authormark{1,7}
}

\address{
\authormark{1}John A. Paulson School of Engineering and Applied Sciences, Harvard University, 29 Oxford Street, Cambridge, MA 02138, USA\\
\authormark{2}Division of Physics, Mathematics and Astronomy, and Alliance for Quantum Technologies (AQT), California Institute of Technology, 1200 E. California Blvd., Pasadena, California 91125, USA\\
\authormark{3}Department of Physics, University of Texas at Austin, Austin, Texas 78712, USA\\
\authormark{4}Department of Electrical Engineering \& State Key Lab of THz and Millimeter Waves, City University of Hong Kong, Kowloon, Hong Kong, China\\
\authormark{5}HyperLight Corporation, 501 Massachusetts Avenue, Cambridge, MA, 02139, USA\\
\authormark{6}shaolb@seas.harvard.edu\\
\authormark{7}loncar@seas.harvard.edu\\
}


\begin{abstract}
We demonstrate conversion of up to 4.5 GHz-frequency microwaves to 1500 nm-wavelength light using optomechanical interactions on suspended thin-film lithium niobate.
Our method utilizes an interdigital transducer that drives a free-standing 100 $\mu$m-long thin-film acoustic resonator to modulate light travelling in a Mach-Zehnder interferometer or racetrack cavity.
Owing to the strong microwave-to-acoustic coupling offered by the transducer in conjunction with the strong photoelastic, piezoelectric, and electro-optic effects of lithium niobate, we achieve a half-wave voltage of $V_\pi = 4.6$ V and $V_\pi = 0.77$ V for the Mach-Zehnder interferometer and racetrack resonator, respectively.
The acousto-optic racetrack cavity exhibits an optomechancial single-photon coupling strength of 1.1 kHz.
Our integrated nanophotonic platform coherently leverages the compelling properties of lithium niobate to achieve microwave-to-optical transduction. 
To highlight the versatility of our system, we also demonstrate a lossless microwave photonic link, which refers to a 0 dB microwave power transmission over an optical channel.
\end{abstract}

%%%%%%%%%%%%%%%%%%%%%%%%%%  body  %%%%%%%%%%%%%%%%%%%%%%%%%%
\section{Introduction}
Conversion of information between the microwave and optical domains is a key ingredient for classical and quantum hybrid signal processing, computing, and networking \cite{Marpaung2019NP, Wendin2017RPP, Rueda2016Optica, Stannigel2010prl, Tsang2010pra}.
Among the many approaches to achieve coherent quantum transduction, electrically-coupled optomechanical systems have emerged as a promising candidate \cite{Schuetz2015prx}. 
Experimental progress includes suspended structures such as optical waveguides in microwave cavities \cite{Fan2019np}, membranes in free-space Fabry-P{\'e}rot cavities \cite{Higginbotham2018np, Andrews2014np, Bagci2014nature}, and nanoscale piezoelectric optomechanical crystals (OMCs) \cite{Forsch2019Arxiv, Jiang2019arxiv, Balram2016np, Vainsencher2016APL, Bochmann2013NP, Fan2013apl}.
While optical waveguides in bulk microwave cavities benefit from the high quality ($Q$) factors of microwave resonance, the suspended membranes achieve a high photon number conversion efficiency leveraging a triple resonance of microwave, mechanical, and optical fields. 
Large-scale integration of these devices is, however, challenging and has not been demonstrated yet. 
On the other hand, OMCs \cite{MacCabe2019arxiv, Liang2017Optica, Balram2014optica, Fang2016np,  Burek2016optica, Eichenfield2009Nature} provide a fully-integrated platform featuring gigahertz mechanical frequencies, and megahertz optomechanical coupling strengths, while limitations due to surface effects are becoming more well-understood \cite{MacCabe2019arxiv}. 
Microwave-to-mechanical (i.e. electromechanical) coupling to OMCs has been achieved using piezoelectric materials, such as aluminum nitride \cite{Vainsencher2016APL, Bochmann2013NP, Fan2013apl}, gallium arsenide \cite{Forsch2019Arxiv, Balram2016np}, and lithium niobate (LN) \cite{Jiang2019arxiv}.
However, the demonstrated electromechanical couplings are inefficient due to the mismatch between the mechanical resonant modes to microwaves \cite{Jiang2019arxiv, Bochmann2013NP, Fan2013apl} or travelling mechanical waves \cite{Forsch2019Arxiv, Balram2016np, Vainsencher2016APL}.

To address the weak microwave-to-mechanical conversion of current integrated devices, we use free-standing LN thin-film acoustic (i.e.~mechanical) resonators with low-loss optical resonators.
Using an interdigital transducer (IDT), our acousto-optic devices demonstrate up to 50\% coupling efficiency from microwave inputs to acoustic resonator modes, thereby enabling efficient optical modulation using Mach-Zehnder interferometers (MZI) and racetrack cavities.
Our approach benefits from the strong piezeoelectricity and electro-optic effects of LN \cite{Lejman2016NatComm, Andrushchak2009jap, Weis1985apa} in conjunction with the photoelastic effect to achieve microwave-to-optical conversion.
Specifically, using a 100 $\mu$m long optical waveguide embeded within a 3.33 GHz acoustic resonator with a $Q$ factor of 3,600, our MZI exhibits a low half-wave voltage $V_\pi$ of 4.6 V.
Moreover,  the half-wave-voltage-length product $V_\pi L$, the figure of merit for optical modulators, is as low as 0.046 V$\cdot$cm, which is a 50-fold reduction over the start-of-the-art electro-optic modulators \cite{Wang2018nature}. 
This comes at expense of reduced microwave-to-optical conversion bandwidth of around 1 MHz, which is significantly smaller than that of electro-optic MZI approaches \cite{Wang2018nature}, nonetheless it is much greater than what has previously been demonstrated using using a microwave, mechanical, and optical triple resonance \cite{Higginbotham2018np}.
Our racetrack cavity features an optical $Q$ factor of over $2\times10^6$, thereby enabling single optical sideband conversion with an effective $V_\pi$ of 0.77 V, a photon number conversion efficiency of 0.0017 \% for an optical power of 1 mW, and an acousto-optic (i.e optomechanical) coupling strength of 1.1 kHz. 
Though this acousto-optic coupling strength is lower than that of state-of-the-art OMCs, the overall microwave-to-optical conversion efficiency is improved due to the enhanced microwave-to-mechanical coupling.
Finally, to illustrate its efficient microwave-to-optical conversion, we demonstrate a loss-less microwave-photonic link with a $\sim$50 mW optical power routing on chip. 

\section{Device design and fabrication}
We utilize IDTs to drive our acoustic resonators due to their efficient electromechanical coupling and ease of fabrication. 
Notably, IDTs are widely used in electro-acoustic signal processing at up to hypersonic (greater than 1 GHz) frequencies \cite{Campbell1989book}. 
Furthermore, they have been used in optical applications to diffract guided beams \cite{Tsai1990book}, modulate cavities \cite{Tadesse2014nc}, drive photonic molecules \cite{Kapfinger2015nc}, and even break time-reversal symmetry \cite{Sohn2018np}.
Therefore, integration of IDT-coupled acoustic resonators \cite{Shao2019arxiv, Yang2018IEEE} with high performance optical devices fabricated in LN \cite{Zhang2017optica, Desiatov2019optica} offers the possibility for efficient acoustically-medicated microwave-to-optical conversion.

\begin{figure*}[ht]
    \centering
    \includegraphics[center]{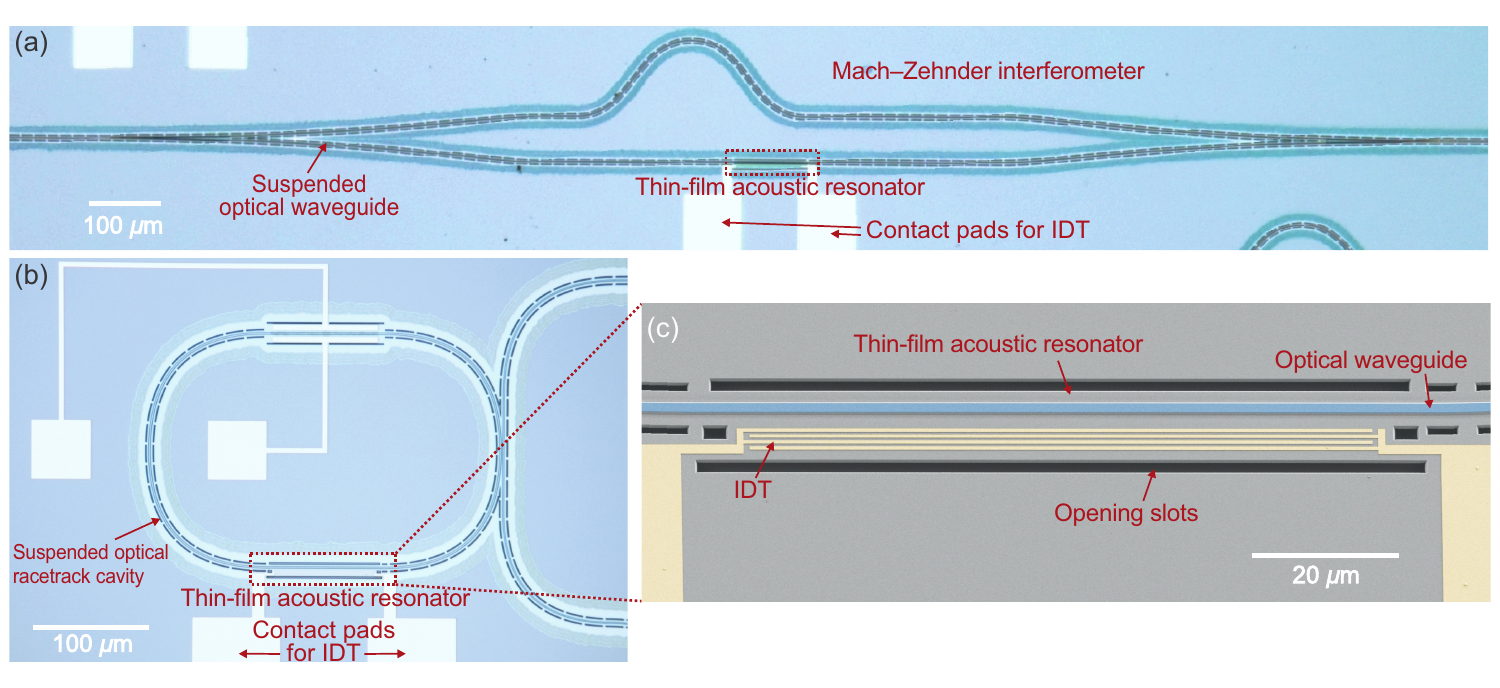}
    \caption{Integrated acousto-optic devices on suspended thin-film LN. (a) Microscope image of a suspended acousto-optic MZI. The interferometer is unbalanced to allow phase control by laser detuning. (b) Microscope image of a suspended optical racetrack cavity with a thin-film acoustic resonator. The suspended regions adjacent to the optical waveguide are identified by a different color, which is darker than the plain substrate in (a) and lighter in (b).
    (c) False color scanning electron microscope image of the acoustic resonator with an IDT and an optical waveguide. 
    }
    \label{fig:schematic}
\end{figure*}

Figures \ref{fig:schematic}(a) and \ref{fig:schematic}(b) shows our acousto-optic devices in the MZI and the racetrack cavity configurations, respectively.
IDT-coupled acoustic resonators host optical waveguides and modulate the phase acquired by the optical signal propagating in one arm of the MZI, which results in intensity modulation of transmitted optical signal  (Fig.~\ref{fig:schematic}(a)). 
Similar effect is responsible for modulating the optical resonance of the racetrack cavity (Fig.~\ref{fig:schematic}(b)).
For our optical components, we employ rib waveguides which are defined by etching a 0.8-$\mu$m-thick X-cut LN thin film to a depth of 0.4 $\mu$m. 
The width of the optical waveguide is 0.95 $\mu$m and 1.3 $\mu$m in the case of MZI and the racetrack cavity, respectively. 
In the case of latter, the larger waveguide cross-section is used to reduce the propagating loss due to sidewall roughness. 
To reduce bending loss and to minimize mode conversion between transverse electric (TE) and transverse magnetic (TM) modes, we employ quadratic B\'ezier curves for our optical waveguides, which is a generalized form of more conventional Euler curves \cite{Cherchi2013OE}.
To confine the acoustic waves, the LN waveguide is released by removing a sacrificial silicon dioxide layer ( Fig.~\ref{fig:schematic}(c)). 
The suspended thin film acoustic resonator is formed by the long slots on both sides of a LN waveguide section.
The length of each acoustic resonator is 10 $\mu$m, and the width is 100 $\mu$m.
The electrode width of 90 $\mu$m is used to match the impedance of IDT with the 50 $\Omega$ impedance of the driving electronics.
Electrode pitch and width of IDTs are controlled to allow excitation of acoustic modes at different frequencies. 
Specifically, an IDT with four electrodes, pitch of 0.6 $\mu$m, and width of 0.3 $\mu$m is used for the acoustic resonator in the MZI configuration. 
This allows coupling to acoustic modes in the range of 1 to 4.5 GHz with peak coupling efficiency occuring around 3 GHz. 
In the racetrack cavity configuration, we use an IDT featuring pitch of 0.86 $\mu$m, width of 0.43 $\mu$m, and same number of electrodes (four). This allows the highest coupling efficiency to acoustic modes around 2 GHz.

To fabricate each device, three layers of electron beam lithography are used to define the LN optical waveguides, the opening slots for the acoustic resonator and the release of the LN layer, and the metal electrodes needed for IDT.
The LN is etched using reactive ion etching, and the fabricated devices feature a sidewall angle of $~70^{\circ}$.
The metal electrodes are defined using a lift-off process:
PMMA resist is patterned as a sacrificial layer, then a 75-nm-thick gold layer with a 8-nm-thick chromium adhesion layer is deposited using electron-beam evaporation, and the device is immersed in a solvent to lift off the resist.
Finally, the release of the LN device from the substrate is achieved using buffered oxide etchant, which removes the underneath sacrificial oxide layer through the completely etched slots of the LN layer.

\section{Description of acoustically-mediated microwave-to-optical conversion}
\begin{figure*}[ht]
    \centering
    \includegraphics[center]{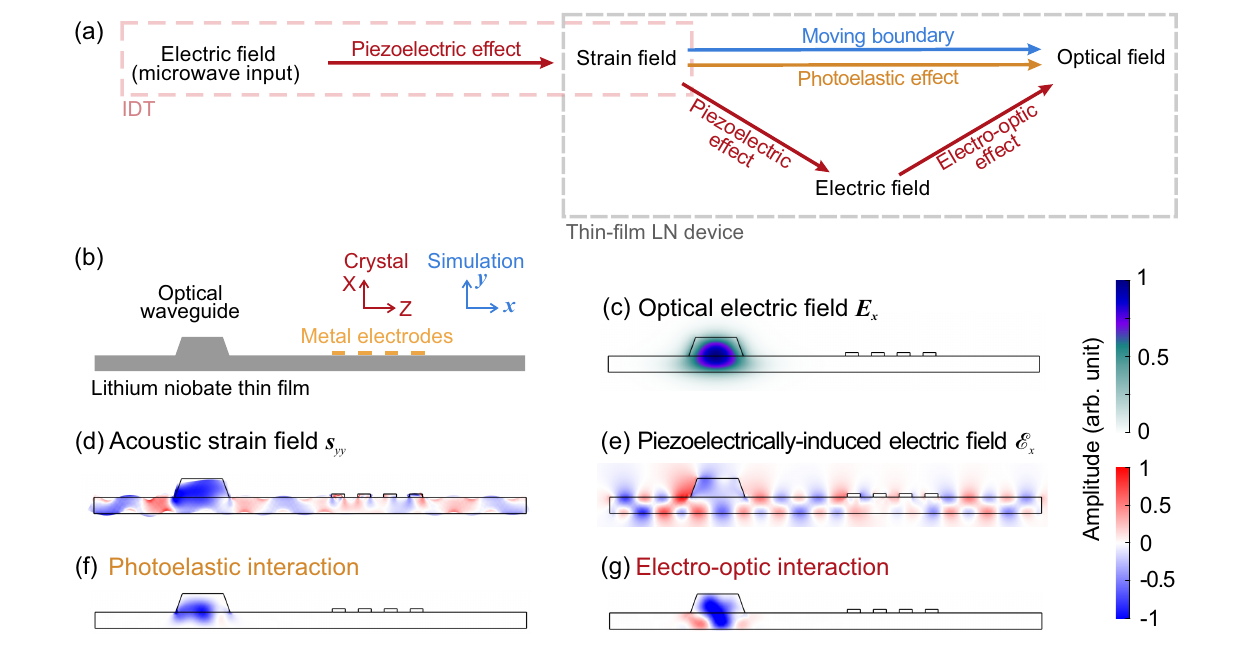}
    \caption{Numerical simulation of the acoustically-mediated microwave-to-optical conversion. 
    (a) Couplings between the microwave, acoustic, and optical fields that facilitate microwave-to-optical conversion.
    (b) Device schematic used for the 2D numerical simulation. The crystal orientation and coordinate system are shown. The top width of the optical waveguide is 0.95 $\mu$m (c) Electric field $E_x$ of the fundamental TE optical mode. (d) $s_{yy}$ component of the acoustic strain field for the 3.24 GHz acoustic mode, and resulting (e) electric fields $\mathcal{E}_x$ induced by piezoelectric effect. We note that $s_{yy}$ has the largest contribution to the photoelastic interaction shown in (f). (g) Electro-optic interactions between the optical TE mode and acoustic fields, mediated by piezo-electric effect. In (f) and (g) the interaction is described by an induced optical refractive index change, calculated by multiplying the optical electric field, the acoustic field, and the interaction matrices. Color scale bars in (d) and (e) are normalized individually, while those in (f) and (g) are the same.}
    \label{fig:simulation}
\end{figure*}

Our acoustic resonator mediates the microwave-to-optical conversion by coupling to the microwave input via the IDT and modulating light due to acoustic modes (Fig.~\ref{fig:simulation}(a)).
The optical modulation is enabled by a generalized acousto-optic interaction that comprises conventional optomechanical couplings of photoelastic and weak moving boundary effects, as well as cascaded piezoelectric and electro-optic effects, which feature a coupling strength comparable to photoelastic alone. 
We perform simulations to understand and engineer the interplay between these three effects in order to maximize the overall acousto-optic interaction. 

The coupling strength of each interaction is evaluated using a 2D numerical model based on the device cross section and crystal orientation shown in  Fig.~\ref{fig:simulation}(b).
To avoid double-counting the coupling strength, we use photoelastic (electro-optic) coefficients under a constant electric field (strain, i.e. clamped) condition following that presented in Ref.~\cite{Weis1985apa, Marculescu1973fr}.
Owing to the LN crystal orientation chosen for our device, the generalized acousto-optic interaction is much stronger for the guided transverse electric (TE) mode (Fig.~\ref{fig:simulation}(b)) than for the transverse magnetic (TM) mode at 1550 nm, due to the strong electro-optic and photoelastic coefficients,  $r_{33}$ and $p_{31}$, respectively, for TE polarization.
The elecro-optic coefficients form a third-order tensor, and $r_{33}$ relates $n_{ZZ}$, the optical index change of the crystal $ZZ$ component  (indicated by the first 3 in the subscript), to $\mathcal{E}_Z$, the electrical field in $Z$ direction (indicated by the first 3 in the subscript).
The photoelastic coefficients form a forth-order tensor, and $p_{31}$ relates $n_{ZZ}$ to $s_{XX}$, the strain of crystal $XX$ component (indicated by the 1 in the subscript). 
Thus the strain $s_{XX}$ (or $s_{yy}$ in simulation coordinate) contributes most to the photoelastic interaction. 
Detailed discussions are provided in the Supplementary Material.
Figures \ref{fig:simulation}(d) and \ref{fig:simulation}(e) plot the simulated acoustic strain $s_{yy}$ and electric field $\mathcal{E}_x$ of a 3.24 GHz mode, both of which have the same sign across the optical waveguide region, thus contributing constructively to the overall modulation to the optical refractive index. 

The contributions of moving boundary, electro-optic and photoelastic effects are calculated by integrating the products of the acoustic and optical field components with the corresponding coupling matrices.
Acousto-optic interaction strengths between various acoustic modes and optical TE or TM mode are summarized in Table S1.
For the 3.24 GHz acoustic mode, Fig.~\ref{fig:simulation}(f) and \ref{fig:simulation}(g) show the induced refractive index changes of the TE mode by the photoelastic and electro-optic effects.
Based on this result, we extract an acousto-optic single-photon coupling strength $g_0 = 1.6$ kHz for the racetrack cavity geometry. 
For the acousto-optic MZI, the half-wave-voltage-length product $V_\pi L$ depends on the the acoustic $Q$ factor and microwave-to-acoustic coupling efficiency, where we use experimental values to avoid over-estimation by the 2D simulation.
A half-wave-voltage-length product $V_\pi L = 0.045$ V$\cdot$cm for the acousto-optic MZI is extracted based on the simulated interaction strength, as well as the acoustic $Q$ of 2,000 and electrical-to-acoustic coupling efficiency of 0.5 (corresponding to a -3 dB dip in the $S_{11}$ spectrum) from typical experimental measurements.
Details are provided in the Sec.~1 of Supplementary Material.

\section{Acousto-optic Mach-Zehnder interferometer}
\begin{figure*}[tb]
    \centering
    \includegraphics[center]{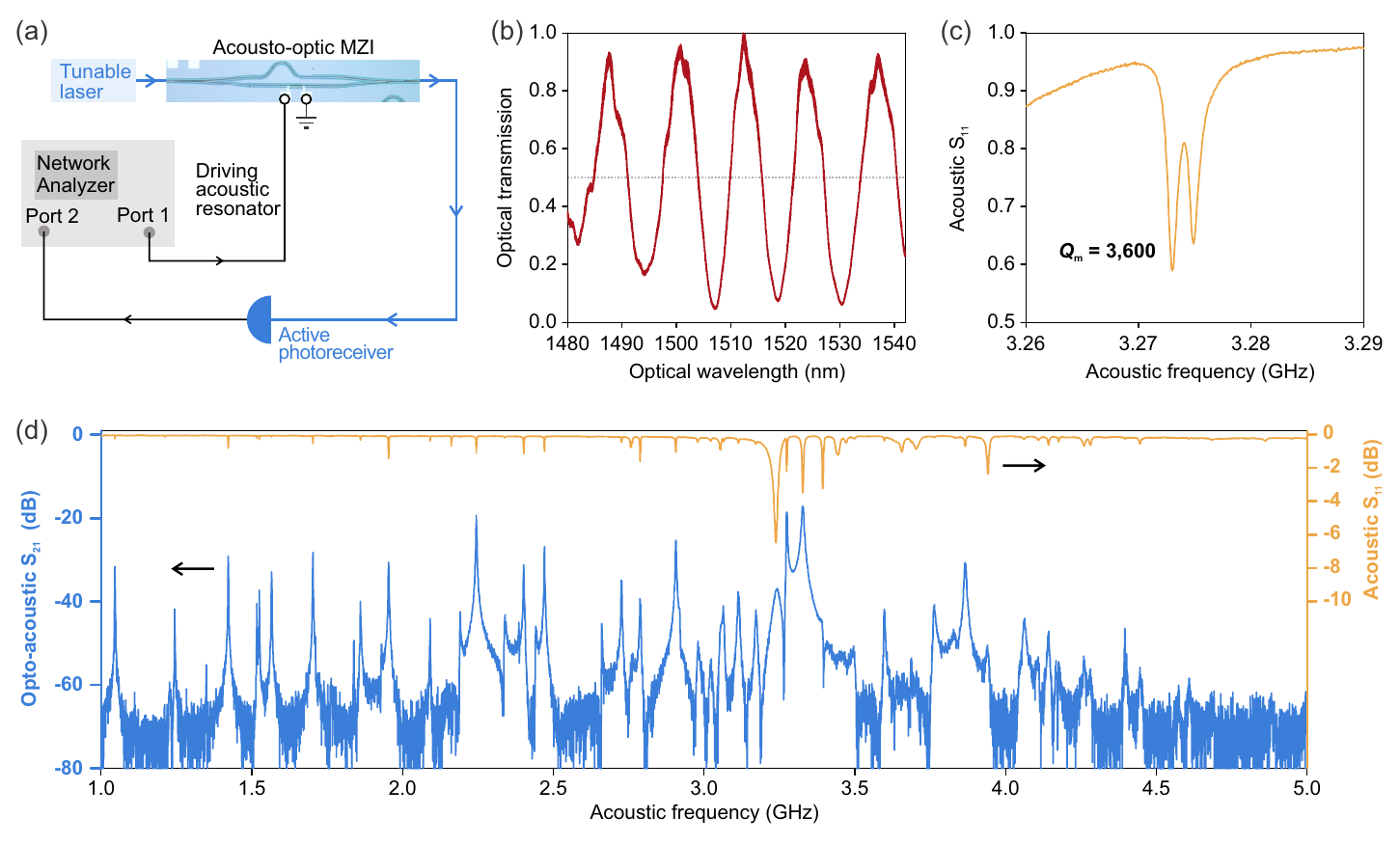}
    \caption{Characterization of the acousto-optic MZI. (a) Simplified experimental schematic for acousto-optic device characterization. (b) Optical transmission of the acousto-optic MZI. (c) $S_{11}$ reflection spectrum of the acoustic resonator. (d) $S_{21}$ spectrum showing an enhanced microwave-to-optical conversion at the resonances indicated by the $S_{11}$ spectrum. 
    The optical power detected by the photoreceiver is 0.25 mW.}
    \label{fig:MZI}
\end{figure*}

We experimentally characterize our acousto-optic MZI using a tunable C-band laser, a vector network analyzer (VNA), and a photoreceiver that features a sensitivity of $\sim$800 V/W (Fig.~\ref{fig:MZI}(a)).
We use lensed fiber to couple light into and out of our structures with a fiber-to-fiber insertion loss of 10 dB ( < 5 dB/facet) for our suspended LN chip.
The periodic variation of optical transmission with wavelength at 10 nm intervals is consistent with the optical path difference in the MZI (Fig.~\ref{fig:MZI}(b)).
To optimize the microwave-to-optical conversion efficiency, the laser wavelength is chosen to be 1534 nm, corresponding to 50\% transmission and indicating a $\pi/2$ phase difference between two optical paths.

We evaluate the acoustic resonances and the microwave-to-acoustic coupling of our devices by measuring the microwave reflection ($S_{11}$) of the IDT.
Our acoustic resonator exhibits multiple resonances in the range between 1.0 and 4.5 GHz  (Fig.~\ref{fig:MZI}(d)).
To correlate measured acoustic modes with that of our the simulations, the acoustic electric field profiles are experimentally-measured using transmission-mode microwave impedance microscopy \cite{Zheng2018pnas, Zheng2018prapplied}, see Sec.~4 in Supplementary Material. 
We measure acoustic $Q$ factors of up to 3,600 (Fig.~\ref{fig:MZI}(c)), similar to the LN OMC devices \cite{Jiang2019arxiv, Liang2017Optica}, and our resonance at the microwave frequency of 3.273 GHz yields a state-of-the-art frequency-quality-factor product of $f Q > 10^{13}$ at room temperature \cite{Yang2018IEEE}.

We characterize the optical modulation induced by the acoustic resonance by the opto-acoustic $S_{21}$ spectrum, where the driving port 1 of the VNA is connected to the IDT of the acoustic resonator and the detecting port 2 is connected to the photoreceiver  [Fig.~\ref{fig:MZI}(a)]. 
The $S_{21}$ spectrum shown in Fig.~\ref{fig:MZI}(d) features several peaks indicating enhanced microwave-to-optical conversion at acoustic resonances, with the strongest responses measured at the 2.24 and 3.33 GHz acoustic modes, in agreement with our simulations (Table S1).
The microwave-to-optical conversion efficiency indicated by the $S_{21}$ depends on both the acoustic $Q$ factor and the overlap between the acoustic mode and the optical mode. 
We extract the half-wave voltage $V_\pi$ of our acousto-optic MZI from the experimental measurements of $S_{21}$ spectrum.  
Under the conditions of our measurement, the MZI half-wave voltage $V_\pi$ is related to the $S_{21}$ by
\begin{equation}
    S_{21} = \left(\frac{\pi\,R_\mathrm{PD}\,I_\mathrm{rec}}{V_\pi}\right)^2,
\end{equation}
in which $R_\mathrm{PD}$ ($I_\mathrm{rec} = 0.25\text{ mW}$) is the sensitivity of (optical power at) the photoreceiver, with derivation given in Sec.~2 of Supplementary Materials.
We find $V_\pi = 4.6$ V (5.8 V) using $S_{21} = -17.4$ dB (-19.3 dB) at the resonance frequency of 3.33 GHz (2.24 GHz), and due to the 100 $\mu$m length of our acoustic resonator, we obtain $V_\pi L = 0.046$ V$\cdot$cm (0.058 V$\cdot$cm), agreeing with that predicted by our simulation.

\section{Acousto-optic racetrack cavity}
\begin{figure*}[tb]
    \centering
    \includegraphics[center]{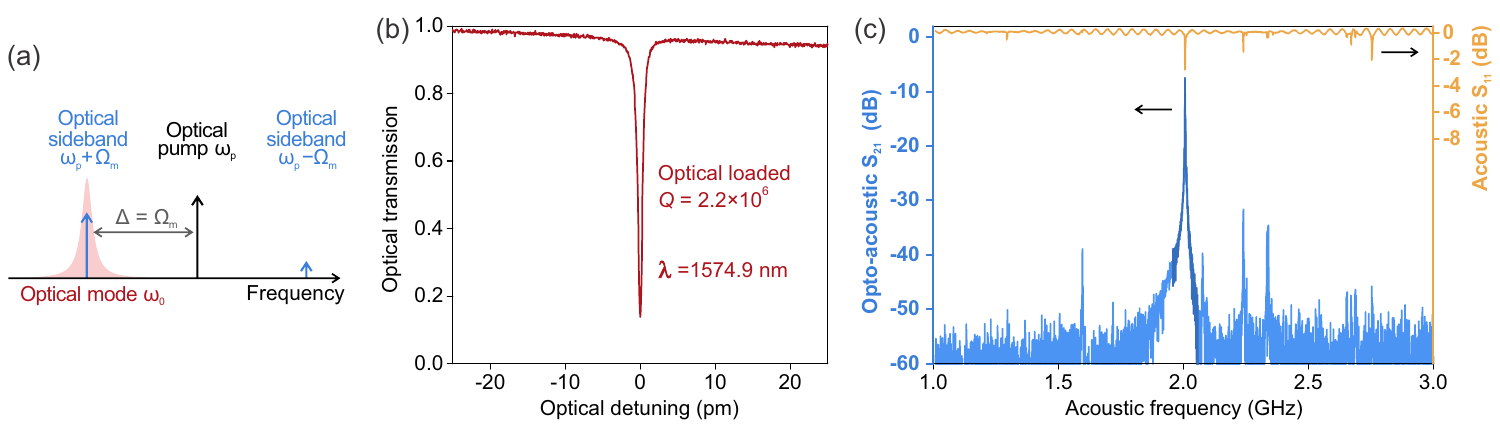}
    \caption{Characterization of the acousto-optic racetrack cavity. 
    (a) Illustration of single-sideband microwave-to-optical conversion using an acousto-optic cavity.  
    (b) Transmission spectrum of an high-$Q$ optical resonance.
    (c) Acoustic $S_{11}$ and opto-acoustic $S_{21}$ spectra. A high resolution measurement around 2 GHz is shown in dark blue. 
    The optical pump wavelength is set to maximize the power received at the photoreceiver.
    }
    \label{fig:AOcavity}
\end{figure*}

Compared with MZI, our racetrack cavity features loaded optical $Q$ factor of $2.2\times 10^6 $ for TE-polarized light of wavelength 1574.9 nm, corresponding to a linewidth of 95 MHz (Fig.~\ref{fig:AOcavity}(b)), allowing us to operate in the microwave sideband-resolved regime.
With the laser blue-detuned by the acoustic resonant frequency from the optical resonance, we generate an optical sideband by the acoustic resonant mode and enhance it using the racetrack resonator (Fig.~\ref{fig:AOcavity}(a)). 
Consequently, we observe a high $S_{21} = -7.5$ dB at the acoustic resonant frequency of 2 GHz with a total optical power of $I_\mathrm{rec} =$ 0.13 mW measured at the photoreceiver (Fig.~\ref{fig:AOcavity}(c)).
We note that the $S_{21}$ quadratically depend on the optical power, and we should consider $S_{21}$ at the same optical power level to compare the conversion efficiency. 
The $S_{21}$ of the racetrack cavity thus results in a much lower effective $V_\pi$ of 0.77 V than the MZI, as determined by the small signal response of a intensity modulator (see Sec.~2 of Supplementary Material).

Next, we determine the overall acousto-optic single-photon coupling strength $g_0$.
In the sideband-resolved regime ($\Omega_m \gg \kappa$) and for weak microwave inputs, the relation between the $S_{21}$ and $g_0$ is given by
\begin{equation}
    S_{21} = \frac{8 g_0^2 \gamma_e \kappa_e^2 R_{PD}^2 I_\mathrm{rec}^2 }{\hbar \gamma^2 \Omega_m^3 \kappa^2 R_\mathrm{load} },
\label{equ:CavityS21}
\end{equation}
where $\kappa$ ($\gamma$) and $\kappa_e$ ($\gamma_e$) are the total loss and external coupling rate of the optical (acoustic) mode, respectively. 
$\Omega_m$ is the frequency of the acoustic mode, and $R_\mathrm{load} = 50\; \Omega$ is the impedance of the input microwave source.
Eq.~\ref{equ:CavityS21} is derived from the equation of motion for the dynamics of the acousto-optic cavity (see Sec.~3 in Supplementary Material),
We estimate the acousto-optic single-photon coupling strength to be $g_0 \sim 1.1$ kHz between the 2.17 GHz acoustic mode and the fundamental TE optical mode, which is in good agreement with our theoretical predictions (see Table S1). 

Another important figure of merit is the photon number conversion efficiency $\eta$ from the microwave frequency to the optical sideband frequency, and it describes the device performance at the single photon level. 
From the derivation described in Sec.~3 of the Supplementary Material, the photon number conversion efficiency $\eta$ is given by
\begin{equation}
    \eta = C_0 \cdot n_\mathrm{cav} \cdot \frac{2\gamma_e}{\gamma}\cdot \frac{2 \kappa_e}{\kappa},
    \label{equ:cvteff}
\end{equation}
where $C_0 = {4 g_0^2}/(\gamma\kappa)$ is the single-photon cooperativity, $n_\mathrm{cav} = {\kappa_e I_\mathrm{opt}}/(\Omega_m^2 \hbar \omega_0)$ is the intracavity optical photon number for the blue-detuned pump light, ${2 \kappa_e}/{\kappa}$ (${2\gamma_e}/{\gamma}$) describes the external coupling efficiency of optical (acoustic) mode. 
Based on the experimentally extracted rates (Table S2), our acousto-optic cavity features an single-photon cooperativity $C_0 = 4\times10^{-8}$, and an photon number conversion efficiency $\eta = $ 0.0017 \% for an optical power of $I_\mathrm{opt} = $  1 mW.
This efficiency could be further improved by acoustic and photonic engineering as discussed later in Sec.~\ref{sec:con}. 

As shown in Eq.~\ref{equ:cvteff}, the photon number conversion efficiency depends on both optomechanical cooperativity ($C_0$) and the microwave-to-mechanical coupling strength (described by $ 2\gamma_e / \gamma$).
Recent progress of LN OMCs \cite{Jiang2019arxiv} has demonstrated unitary optomechanical cooperativity, but their microwave-to-mechanical coupling strength is as low as $10^{-8}$, which would limit the overall photon number conversion efficiency. 
Benefiting from our up to 50\% microwave-to-acoustic coupling to the thin-film acousto-optic resonator, this photon number conversion efficiency could be greater than the OMCs, even though the acousto-optic coupling strength is weaker than that of the OMCs.

\section{Demonstration of a microwave-photonic link}
\begin{figure*}[tb]
    \centering
    \includegraphics[center]{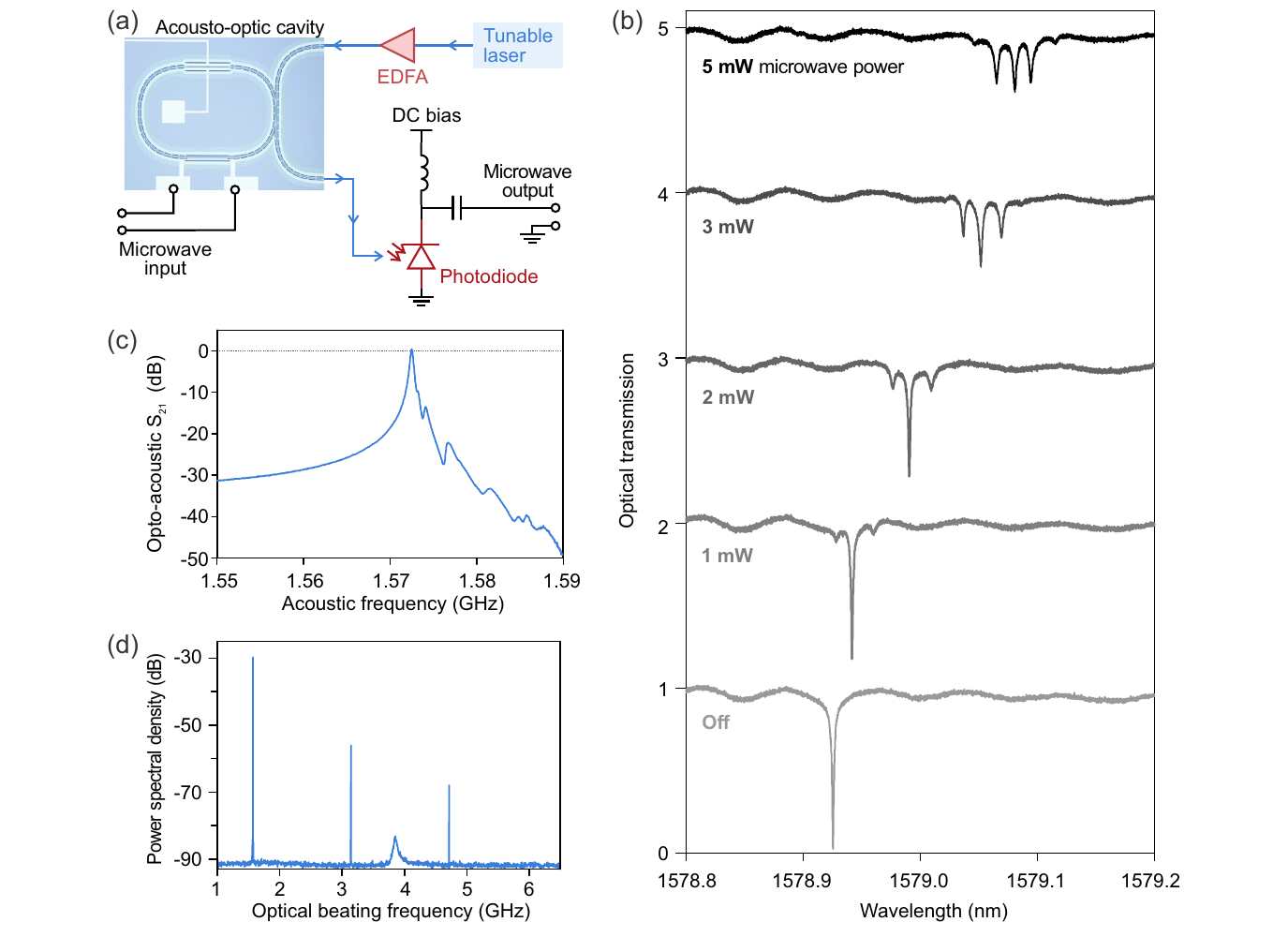}
    \caption{Demonstration of a microwave-photonic link. (a) Experimental schematic. EDFA: Erbium-doped fiber amplifier. 
    (b) Optical transmission of the racetrack cavity for different microwave powers. 
    (c) $S_{21}$ spectrum features a peak microwave power transmission of $\sim 0$ dB. 
    The optical power received at the photodiode $I_\mathrm{rec}$ is 50 mW.
    (d) Microwave spectrum of the photodiode output signal with a  microwave power of 5 mW applied to the IDT of the thin-film acoustic resonator. The laser is blue-detuned from the optical mode by the acoustic resonant frequency $\Omega_m$
    }
    \label{fig:highpower}
\end{figure*}

A microwave photonic link enables low-loss long haul transport and flexible manipulation of microwave signals using optical devices by up-converting microwave frequencies to optical frequencies.
To benchmark our acousto-optic racetrack device, we demonstrate a narrow-band microwave-photonic link using our acousto-optic racetrack cavity, and a link gain of 0 dB is achieved without the need of an optical amplifier within the link (after the modulation of our acousto-optic device), which would significantly increases the noise of the link. 
The optical pump light is amplified to $\sim$500 mW by an erbium-doped fiber amplifier and is blue-detuned by the acoustic resonant frequency $\Omega_m$ from the optical resonance (Fig.~\ref{fig:highpower}(a)).
We estimate that $\sim 150$ mW of optical power is coupled into the suspended LN waveguide, resulting in $\sim 50$ mW reaching the photodiode.
Importantly, no damage to the waveguide is observed, indicating the ability of our suspended thin-film LN devices to handle large optical powers.
A high power photodiode, with responsivity $R_\mathrm{PD}$ of 0.55 A/W (corresponding to a quantum efficiency of 40\%), is used to detect and down-convert the optical signal that we generate from our our racetrack acousto-optic transducer back to the microwave domain.
With these parameters, we measure an overall microwave link gain of 0 dB at 1.572 GHz (Fig.~\ref{fig:highpower}(c)) for a small microwave input signal at -20 dBm. 
By reducing the fiber-to-chip coupling loss to the previously-demonstrated value of 1.7 dB/facet \cite{He2019OL}, a microwave link with gain of 6.6 dB may be achieved in principle, with the possibility of a gain of at least 9 dB if tapered fibers are employed \cite{Burek2017PRApplied, Sipahigil2016Science}.

Next, we characterize the response at higher powers of microwave input. 
With increasing microwave powers at the acoustic resonant frequency $\Omega_m\,\sim$2 GHz, optical sideband dips are observed in the transmission spectra (Fig.~\ref{fig:highpower}(b)), which agree with theoretical predictions (Fig.~S3). 
The red-shift of the optical resonance with increasing input microwave powers results from the heating of the acoustic resonator. 
As a result of efficient microwave-to-optical conversion, a pair of second order sideband dips can be  observed in the optical transmission spectrum with only 5 mW of microwave input power.

Parking the laser at $\Omega_m$ detuning from the optical mode, up to 3rd order harmonic signals are observed at the photodiode output with a microwave power of 5 mW (Fig.~\ref{fig:highpower}(d)). 
The additional broad peak observed at 3.85 GHz is a result of the suspended optical racetrack cavity since it is only observed when the pump laser is close to the optical resonance. 
We speculate that the acoustic mode along the suspended optical waveguide causes this additional peak by spontaneous Brillouin scattering, as our suspended racetrack cavity has the similar geometry with that in Ref.~\cite{Otterstrom2018science}. 

\section{Conclusions and outlook}
\label{sec:con}
We demonstrate an integrated acousto-optic platform on thin film LN, which converts acoustic waves in microwave domain to optical light by a generalized acousto-optic interaction. 
Efficient microwave-to-acoustic coupling has been achieved using our IDT-coupled LN thin-film acoustic resonator.
This addresses the coupling issue of current OMC-based platforms using mechanically-mediated microwave-to-optical converters. 
To further improve the photon number conversion efficiency, a variety of efforts in acoustic and photonic engineering can be made. 
For example, the acoustic resonator can be operated under vacuum and cryogenic environments to achieve higher $Q$ factors, and the clamping loss of the suspended structures can be further reduced using phononic crystals \cite{Fang2016np}.
Optical cavities defined by photonic crystal mirrors could improve the acousto-optic coupling strength $g_0$ with larger overlap with the acoustic resonator, while the presented racetrack cavity only partially sits in the acoustic resonator. 
Another order-of-magnitude improvement could be obtained by bringing both the pump light and generated optical sideband into resonance \cite{Rueda2016Optica}. 
With a double optical resonance, which can be found in coupled cavities or due to scattering in a single ring cavity, the term $\Omega^2$ ($\sim$ GHz) in the denominator of intracavity optical photon number $n_\mathrm{cav}$ (Eq.~\ref{equ:cvteff}) is replaced by the optical cavity loss $\kappa^2$ ($\sim$ 10s of MHz) and could result in an improvement of 4 orders of magnitude.

Beyond microwave-to-optical conversion, our acousto-optic platform could also find applications in gigahertz frequency optical comb generation, on-chip optical routing and optical mode conversion.
In these applications, the acoustic resonator could strongly enhance the signal in microwave domain and allow low microwave power operations. 
Compared to microwave electromagnetic resonators, the high acoustic $fQ$ product and smaller acoustic mode volume of our on-chip acousto-optic resonator could enable quantum optomechanics at room temperatures with smaller footprints.

\section*{Acknowledgment}
This work is supported by the STC Center for Integrated Quantum Materials, NSF Grand No.~DMR-1231319, NSF E2CDA Grand No.~ECCS-1740296, NSF CQIS Grand No.~ECCS-1810233, ONR MURI Grant No.~N00014-15-1-2761, and NSF Grant No.~DMR-1707372. N.S.~acknowledges the support of the Natural Sciences and Engineering Research Council of Canada (NSERC) and the AQT Intelligent Quantum Networks and Technologies (INQNET) research program.

\bibliography{LNAOreference}

\pagebreak

%%%%%%%%%%%%%%%%%%%%%%%%%%% SI %%%%%%%%%%%%%%%%%%%%%
{\Large \centering \textbf{Supplementary Material}}
\vspace{10pt}

\renewcommand{\thefigure}{S\arabic{figure}}
\renewcommand{\thetable}{S\arabic{table}}
\renewcommand{\theequation}{S\arabic{equation}}
\setcounter{section}{0}
\setcounter{figure}{0}
\setcounter{table}{0}
\setcounter{equation}{0}

\section{Numerical simulation of the acousto-optic interaction}
We perform a 2D numerical simulation of our device cross-section (Fig.~\ref{fig:opricalMode}(a)) using COMSOL Multiphysics.
Optical and acoustic modes are simulated independently and the acousto-optic interactions are then calculated by the integral of acoustic and optical fields using corresponding nonlinear coefficient matrices.  

\subsection{Simulation of optical and acoustic modes}
\begin{figure}[ht]
    \centering
    \includegraphics{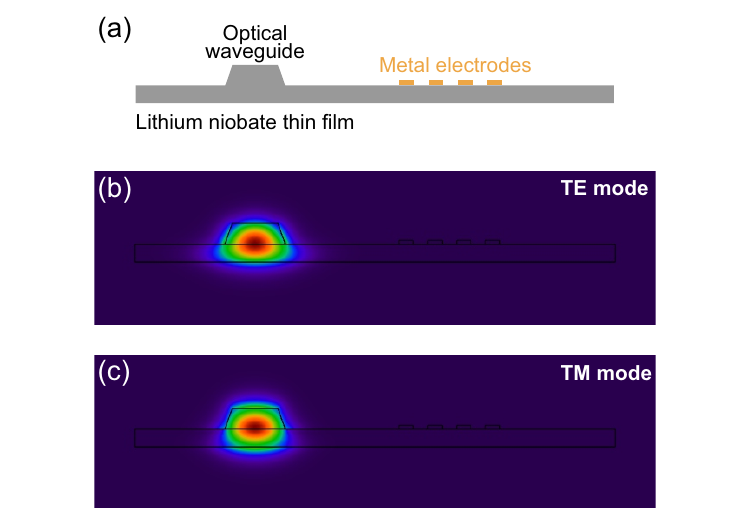}
    \caption{(a) Device structure for 2D numerical simulation. (b), (c) Optical electric field of the fundamental TE and TM modes, respectively.}
    \label{fig:opricalMode}
\end{figure}

The single-mode optical waveguide of our device supports fundamental TE and TM modes (Fig.~\ref{fig:opricalMode}). 
The electric field profiles of the optical modes are used in the calculation of the acousto-optic interaction. 

\begin{figure*}
    \centering
    \includegraphics{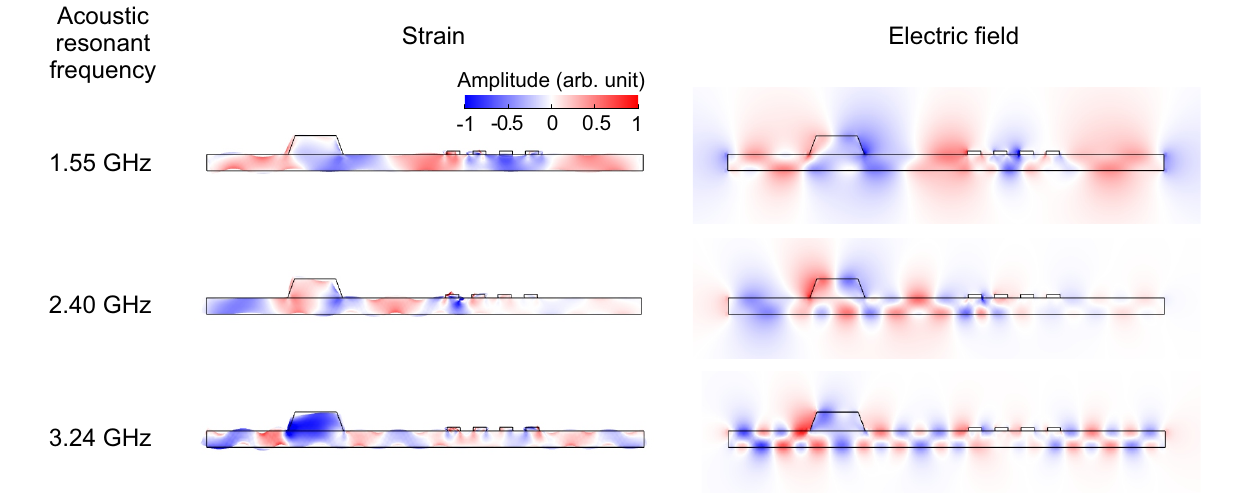}
    \caption{Strain and electric field of three simulated acoustic modes with resonant frequencies of 1.55, 2.40, and 3.24 GHz. The color map is independently normalized for each simulation.}
    \label{fig:acoustiModes}
\end{figure*}

The simulation of the acoustic mode includes strain, electric field, and the piezoelectric effect. 
Multiple acoustic modes with gigahertz resonant frequencies are found in the eigenmode simulation. 
We plot only a few acoustic modes in Fig.~\ref{fig:acoustiModes}.
The electrical excitation of these acoustic modes are enabled by the interdigital transducers (IDTs). 

\subsection{Calculation of acousto-optic interactions}
The acousto-optic interactions are calculated by integrating the optical and acoustic modes with matrices that describe moving boundary, photoelastic and electro-optic effects. 
Calculations here are based on theory formulated in previous works \cite{Burek2016optica, Balram2014optica, Eichenfield2009Nature, Newnham2005book}.

In our work, the acousto-optic interactions are described by the change of optical mode index due to the acoustic mode. 
The acoustic mode amplitude $\alpha$, defined by the maximum displacement, is normalized to a single phonon occupation of the acoustic resonator using $ \hbar \Omega = \frac{1}{2}m_\mathrm{eff}\Omega^2 \alpha^2$, where $\Omega$ is the acoustic frequency.
The effective mass $m_\mathrm{eff}$ of the acoustic mode is given by
\begin{equation}
    m_\mathrm{eff} = L_a \int_D \rho \, Q(\bm{r})^2  d\bm{r} \big/ \max_D \left(Q(\mathbf{r})^2\right),
\end{equation}
where $D$ defines the 2D simulation domain and the coordinate variable $\bm{r} \in D$.  
$L_a$ is the length (perpendicular to the simulation cross-section) of the acoustic resonator, $\rho$ is material mass density, and $Q$ is the displacement field.

The electric field of the optical mode is denoted as $E$, $s$ refers to strain, and $\mathcal{E}$ refers to the electric field of the acoustic mode. 
The mode index modulated by the moving boundary effect is given by 
\begin{equation}
    \Delta n_{0, \mathrm{MB}} = -\frac{n}{2} \frac{\oint \left( Q \cdot \mathbf{\hat n} \right) \left( E_\|^*\Delta \epsilon E_\| -D_\perp^* \Delta \epsilon^{-1} D_\perp \right) dS}{\int E^*\epsilon E dr},
\end{equation}
where $n$ is the optical mode index, $\mathbf{\hat n}$ is the normal vector of the boundary facing outward, and $D$ is the electric displacement field of the optical mode.
The subscripts $\|$ and $\perp$ indicate the parallel and perpendicular components to the boundary. 
The permittivity for the optical electric field is denoted as $\epsilon$, while $\Delta \epsilon = \epsilon_\mathrm{LN} - \epsilon_\mathrm{air}$, and $\Delta \epsilon^{-1} = \epsilon_\mathrm{LN}^{-1} - \epsilon_\mathrm{air}^{-1}$. 

The mode index modulated by the photoelastic effect is given by 
\begin{equation}
\label{equ:deltaNdB}
    \Delta n_{0, \mathrm{PE}} = \frac{\epsilon_0 n^5}{2} \frac{ \int dr \begin{pmatrix} 
E_x^* & E_y^* & E_z^* 
\end{pmatrix}
\begin{pmatrix} 
dB_1 & dB_6 & dB_5 \\
dB_6 & dB_2 & dB_4 \\
dB_5 & dB_4 & dB_3
\end{pmatrix}
\begin{pmatrix} 
E_x \\
E_y \\
E_z 
\end{pmatrix}
}{\int E^* \epsilon E dr},
\end{equation}
where $\epsilon_0$ is the vacuum permittivity, and $B_k$ $(k=1-6)$ is the optical indicatrix.
The changes of indicatrix coefficient $dB_k$ $(k=1-6)$ due to the strain $s_k\; (k=1-6)$ is given by 
\begin{equation}
    \begin{pmatrix} 
dB_1 \\ dB_2 \\ dB_3 \\ dB_4 \\ dB_5 \\ dB_6
\end{pmatrix}
=
\begin{pmatrix}
p_{33} & p_{31} & p_{31} & 0 & 0 & 0 \\
p_{13} & p_{11} & p_{12} & 0 & p_{14} & 0 \\
p_{13} & p_{12} & p_{11} & 0 & -p_{14} & 0 \\
0      &    0   &    0   & p_{66} & 0 & p_{14} \\
0      & p_{41} & -p_{41} & 0 & p_{44} & 0 \\
0      &    0   &    0   & p_{41} & 0 & p_{44} 
\end{pmatrix} 
\begin{pmatrix} 
s_1 \\ s_2 \\ s_3 \\ s_4 \\ s_5 \\ s_6
\end{pmatrix}.
\end{equation}
where $p_{jk}$ are the primary elasto-optic coefficients in the condition of a constant electric field for lithium niobate (LN), where the secondary effect via piezoelectricity and electro-optics is excluded \cite{Weis1985apa, Marculescu1973fr}. 
The photoelastic matrix is rotated according to the crystal orientation in our device -- 
 X-cut thin-film LN with acoustic wave propagating in the Z direction of the crystal. 
The coordinate representations for the simulation and crystal are shown in Fig.~2(a).

The mode index modulated by the electro-optic effect $\Delta n_{0, \mathrm{EO}}$ is of the same form of Eq.~\ref{equ:deltaNdB}, with the changes of indicatrix coefficients \cite{Newnham2005book} given by
\begin{equation}
    \begin{pmatrix} 
dB_1 \\ dB_2 \\ dB_3 \\ dB_4 \\ dB_5 \\ dB_6
\end{pmatrix}
=
\begin{pmatrix}
r_{33} & 0   & 0 \\
r_{13} & 0   & -r_{22} \\
r_{13} & 0   & r_{22} \\
0      & -r_{22}  &    0  \\
0      & 0   & r_{51} \\
0      & -r_{51}   & 0
\end{pmatrix} 
\begin{pmatrix} 
\mathcal{E}_x \\
\mathcal{E}_y \\
\mathcal{E}_z
\end{pmatrix}, 
\end{equation}
where $r_{jk}$ is the primary electro-optic coefficients in the condition of constant strain in which secondary effects via piezoelectricity and photoelasticity is excluded. 
The above matrix is rotated according to the crystal orientation in our device. 

The overall relative refractive index change due to a single phonon is given by,
\begin{equation}
    \Delta n_{0,\mathrm{tot}} = \Delta n_{0,\mathrm{MB}} + \Delta n_{0,\mathrm{PE}} + \Delta n_{0,\mathrm{EO}}.
    \label{equ:delntot}
\end{equation}

\begin{table*}[thb]
    \centering
    \caption{Numerical simulation results of acousto-optic interactions}
    \begin{tabular}{m{1cm}m{1.8cm}m{1.4cm}m{1.4cm}m{1.4cm}m{1.4cm}ccc}
        \hline
        Optical mode & Acoustic mode freq. & $\Delta n_{0,\mathrm{MB}}$  & $\Delta n_{0,\mathrm{PE}}$ & $\Delta n_{0,\mathrm{EO}}$ & $\Delta n_{0,\mathrm{tot}}$ & \multicolumn{2}{c}{MZI} & AO cavity \\
        & & & & & & $V_\pi L$ & $V_\pi$ & $g_0$ \\
        & GHz & $\times 10^{-12}$ & $\times 10^{-12}$ & $\times 10^{-12}$ &  $\times 10^{-12}$ & V$\cdot$cm & V & kHz \\
        \hline
        TE & 1.55 & 0.36 & 1.40 & 30.5 & 32.26 & 0.0692 & 6.92 & 0.5\\
        TE & 2.17 & -0.84  & 1.11 & 71.43 & 71.70 & 0.0436 & 4.36 & 1.1 \\
        TE & 2.40  & -0.80 & -5.08 & 23.06 & 17.17 & 0.2009 & 20.1 & 0.27 \\
        TE & 3.16 & -3.07 & 41.68 & 26.38 & 64.99 & 0.0703 & 7.03 & 1.0 \\ 
        TE & 3.24 & -4.24 & 47.84 & 58.91 & 102.5 & 0.0454 & 4.54 & 1.6 \\
        TM & 1.55 & 4.19 & 24.29 & -8.79 & 19.69 &  0.113 & 11.3 & 0.3\\
        TM & 2.17 & 8.39 & 67.97 & -24.03 & 52.33 & 0.0599 & 5.99 & 0.8 \\
        TM & 2.40 & 0.87 & 22.26 & -9.37 & 13.72 & 0.2505 & 25.1 & 0.2\\
        TM & 3.16 & 15.09 & 42.06 & -4.28 & 52.87 & 0.0863 & 8.63 & 0.8\\
        TM & 3.24 & 21.03 & 70.63 & -15.75 & 75.90 & 0.0616 & 6.16 & 1.2\\
        \hline
    \end{tabular}
    \label{tab:VpiLpara}
\end{table*}

\subsection{Calculation of $V_\pi L$}
The half-wave-voltage-length product $V_\pi L$, characterizing the modulation efficiency, defines the voltage that is required to achieve a $\pi$ phase shift for a modulation length $L$.
Here, we derive the $V_\pi L$ from the simulated refractive index changes with additional information on $Q$ factors and coupling of the acoustic resonator.
While the overall refractive index change in Eq.~\ref{equ:delntot} quantifies the optical phase shift (or index change) due to a single phonon in the acoustic resonator,
one must relate the in-cavity phonon number to the applied microwave power. 
As discussed later in Sec.~\ref{sec:aoc}, the in-cavity phonon number is given by 
\begin{equation}
\label{equ:Npn}
    N_{pn} = \frac{4 \gamma_e}{\gamma^2}\; N_{in} = \frac{4 \gamma_e}{\gamma^2}\; \frac{P_{in}}{\hbar \Omega_m},
\end{equation}
where $N_{in} ={ P_{in}}\big/{\hbar \Omega_m}$ is the phonon input rate with the resonant frequency $\Omega_m$ of the acoustic mode and input power $P_{in}$.
The decay are and external coupling rates of the acoustic mode is $\gamma$ and $\gamma_e$, respectively. 
Given the input impedance $R_{in} = 50 \Omega$, the relation between input power and peak voltage $V_p$ is given by
\begin{equation}
\label{equ:PinV}
    P_{in} = \frac{1}{2}\;\frac{V_p^2}{R_{in}}.
\end{equation}
The number of in-cavity phonons $N_{pn}$ required for a $\pi$ phase shift is given by,
\begin{equation}
    \frac{2 \pi}{\lambda} \Delta n_{0,\mathrm{tot}} \sqrt{N_{pn}} L = \pi
    \label{equ:piphase}
\end{equation}
where $\lambda$ is the optical wavelength. 
Taking Eqs.~\ref{equ:Npn} and \ref{equ:PinV} in to Eq.~\ref{equ:piphase}, we derive the $V_\pi L$ of the device:
\begin{equation}
    V_\pi L = \frac{\lambda}{ 2 \Delta n_{0,\mathrm{tot}}} 
             \sqrt{\frac{ \gamma^2 \hbar \Omega_m R_{in} }{ 2 \gamma_e}}.
\end{equation}

\subsection{Calculation of acousto-optic single-photon coupling strength $g_0$}
For our thin-film acoustic resonator that is coupled to an optical racetrack cavity, the acousto-optic single-photon coupling strength $g_0$ can be derived using the 2D simulation results using
\begin{equation}
    g_0 = \omega_0 \eta_\mathrm{cav}\,\frac{\Delta n_{0,\mathrm{tot}}}{n},
\end{equation}
where $\omega_0$ is the optical resonant frequency, and $\eta_\mathrm{cav}$ is the ratio of waveguide length in the acoustic resonator to that of the racetrack cavity.

\subsection{Estimate $V_\pi L$ and $g_0$ using the numerical simulation results}
We estimate the $V_\pi L$ for the Mach–Zehnder interferometer (MZI) and $g_0$ for the acousto-optic cavity from simulation. 
To be consistent with the experiments, the typical measured acoustic $Q$ factors $Q_m = 2,000$ ($\gamma = \Omega_m / Q_m$) and $\gamma_e/\gamma = 0.15$ (corresponding to a 3 dB dip in $S_{11}$ measurements) are employed in the following calculation.
The length of the acoustic resonator (in direction perpendicular to the simulation cross-section) is $L_a = 100 \;\mu\mathrm{m}$. 
The output impedance of the microwave source is $R_{in} = 50 \;\Omega$.
For the acoustic-optic cavity shown in Fig.~1, the relative length of the optical waveguide in the acoustic resonator is $\eta_\mathrm{cav} = 0.15$. 
Table \ref{tab:VpiLpara} summarizes the interactions between optical modes and acoustic modes.

\section{Derivation of $V_\pi$ from experimental measurements}
\subsection{Acousto-optic Mach-Zehnder interferometer}
Here we relate the half-wave voltage $V_\pi$ to the measured opto-acoustic $S_{21}$ for the acousto-optic MZI. 
The phase modulation of one optical path is given by
\begin{equation}
    E_{p1}(V) = \frac{E_0}{\sqrt{2}} \exp\left( i \pi V/ V_\pi + i \phi_b \right),
\end{equation}
where $E_0$ is the input optical field of the MZI, $\phi_b$ is the bias phase between two optical paths, and $V$ is the applied voltage.
The other optical path of MZI is not modulated, and the optical field is given by $E_{p2}(V) = E_0 / \sqrt{2}$.
The optical field at the output of the MZI is given by
\begin{equation}
\begin{split}
    E_{\mathrm{out}}(V) &= \frac{E_{p1}(V) + E_{p2}(V)}{\sqrt{2}} \\
    &= \frac{E_0}{2}\, \left(1 + \exp\left( i \pi V/ V_\pi + i \phi_b \right) \right).
\end{split}
\end{equation}
The output optical power is thus given by
\begin{equation}
\begin{split}
    I_\mathrm{out}(V) &\propto E_\mathrm{out}^* \, E_\mathrm{out} \\
    &= \frac{|E_0|^2}{2} \left( 1 + \cos\left( \pi V/V_\pi + \phi_b \right)  \right). 
\end{split}
\end{equation}
The optimum microwave to optical conversion occurs at the bias phase $\phi_b = \pi/2$, which corresponds to the output intensity at half maximum.  
Measured using a potodetector, the opto-acoustic $S_{21}$ for small input signal is given by
\begin{equation}
    S_{21} = \left(\frac{\pi R_{PD} \, I_\mathrm{rec}}{V_\pi} \right)^2 ,
    \label{equ:MZMVpiS21}
\end{equation}
where $I_\mathrm{rec}$ is the DC optical power received at the photodetector, and $R_{PD}$ is sensitivity of the photodetector.
Using Eq.~\ref{equ:MZMVpiS21}, we can derive $V_\pi$ of the acousto-optic MZI by the opto-acoustic $S_{21}$ measurements. 

\subsection{Acousto-optic cavity}
Our acousto-optic cavity operates in the sideband resolved regime, that is the frequency of the microwave signals are greater than the decay rate of the optical mode. 
For a weak microwave signal, the optical transmission is thus close to unitary at the optimum conversion wavelength, which corresponds to that detuned from the optical resonance by the microwave frequency.
Phenomenologically, this can be understood as the light being reversibly pumped into, and out of, the optical cavity due to the acoustic modulation.   
Thus, we consider the acousto-optic cavity as an intensity modulator and the relation in Eq.~\ref{equ:MZMVpiS21} is also used to derive the effective $V_\pi$. 

\section{Conversion between microwave, acoustic, and optical fields in acousto-optic cavity}
\label{sec:aoc}

\subsection{Dynamics of acousto-optic cavity}
Here we consider an acousto-optic system with an acoustic resonator driven by a microwave signal through the piezoelectric effect. 
The Heisenberg-Langevin equations of motion for an optical cavity $a$ coupled to an acoustic resonator $b$ are given by
\begin{align}
\dot{a} &= - \left( i \Delta + \frac{\kappa}{2} \right) a - i g_0 a \left( b+b^\dagger \right) + \sqrt{\kappa_e} a_{in} \label{equ:motiona}\\
\dot{b} &= - \left( i \Omega_m + \frac{\gamma}{2} \right) b - i g_0 a^\dagger a + \sqrt{\gamma_e} b_{in}, 
\label{equ:motionb}
\end{align}
where $a$ and $b$ are the annihilation operators of optical and acoustic modes, respectively, 
$g_0$ is the single-photon coupling strength between the optical and acoustic resonators,
$\Delta = \omega_0 - \omega_p$ is the optical detuning with the pump laser frequency $\omega_p$, the optical resonant frequency is $\omega_0$,  
$\kappa = \kappa_i + \kappa_e$ is the loss of optical mode with intrinsic loss $\kappa_i$ and external coupling rate $\kappa_e$, 
$\Omega_m$ is the acoustic resonant frequency,   
$\gamma = \gamma_i + \gamma_e$ is the loss of acoustic mode with intrinsic loss $\gamma_i$ and external coupling rate $\gamma_e$, and $a_{in}$ and $b_{in}$ are the optical and microwave input field, respectively.

To solve the equations of motion, we consider a single frequency microwave driving $b_{in}$ of the acoustic resonator given by
\begin{equation}
    b_{in} = B_{in} e^{-i \Omega_d t}, 
    \label{equ:bin}
\end{equation}
where $\Omega_d$ is the driving frequency, $B_{in}$ is the amplitude of the input field, and the input microwave power is $P_{in} = \hbar \Omega_m \left|B_{in}\right|^2 $.
In the weak optical mode limit, i.e.~$g_0 a^\dagger a \ll \Omega_m$, the optical back action term ($ i g_0 a^\dagger a $ in Eq.~\ref{equ:motionb}) on the acoustic resonator is neglected. 
Taking Eq.~\ref{equ:bin} into Eq.~\ref{equ:motionb}, the acoustic amplitude $b$ is solved using
\begin{equation}
    \begin{split}
        b &= B e^{-i \Omega_d t} \\
        B &= \frac{\sqrt{\gamma_e}}{i \left( \Omega_m - \Omega_d \right) + \frac{\gamma}{2}}B_{in}. 
    \end{split}
    \label{equ:bsol}
\end{equation}
For a resonant microwave drive ($\Omega_m = \Omega_d$), the in-resonator phonon number $N_{pn}$ is related to the input microwave power by
\begin{equation}
\begin{split}
    N_{pn} &= B^2 \\
           &= \frac{ 4 \gamma_e}{\gamma^2} B_{in}^2 \\
           &= \frac{ 4 \gamma_e}{\gamma^2} N_{in} \\
           &= \frac{ 4 \gamma_e}{\gamma^2} \frac{ P_{in}}{\hbar \Omega_m}.  
\end{split}
\end{equation}

Taking Eq.~\ref{equ:bsol} into Eq.~\ref{equ:motiona}, the equation of motion for the optical mode is re-written as
\begin{equation}
    \dot{a} = - \left( i \Delta + \frac{\kappa}{2} \right) a - i G\, 2 \cos\left(\Omega_d t\right) a  + \sqrt{\kappa_e} a_{in},
    \label{equ:optmotion}
\end{equation}
where $G = g_0 B$ is the frequency shift of optical mode due to the  acoustic field that is present.

\subsection{Optical transmission with active acoustic driving}
We numerically solve Eq.~\ref{equ:optmotion} to investigate the optical transmission spectra with various microwave input powers.
We note that the Eq.~\ref{equ:optmotion} assumes a weak optical input and a linear acoustic resonator.
The normalized optical transmission $T$ under a continuous optical pump $a_{in}$ is given by
\begin{equation}
    T = \left| a_{in} - \sqrt{\kappa_e} a \right|^2 \big/ \left| a_{in} \right|^2.
\end{equation}

As the optical mode is being modulated by an acoustic mode at microwave frequency $\Omega_d$, the optical transmission $T$ is expected to associate an oscillation at the same as well as higher order frequencies due to nonlinearity. 
However, in experiment, the optical transmission spectra are captured by a low frequency (10 MHz) data acquisition card, which does not respond to gigahertz frequencies. 
Numerically, we use an average to calculate the quasi-DC component of the optical transmission using 
\begin{equation}
    T_{DC} = \frac{1}{\Delta t} \int_{t_1}^{t_1+\Delta t} d\tau T(\tau),
\end{equation}
where time $t_1$ is set to be greater than the initial stabilization time in numerical calculation of $a$, and the average time window $\Delta t$ is chosen to be the integer periods of the driving signal, i.e.~$N/\Omega$. 

\begin{figure}
    \centering
    \includegraphics{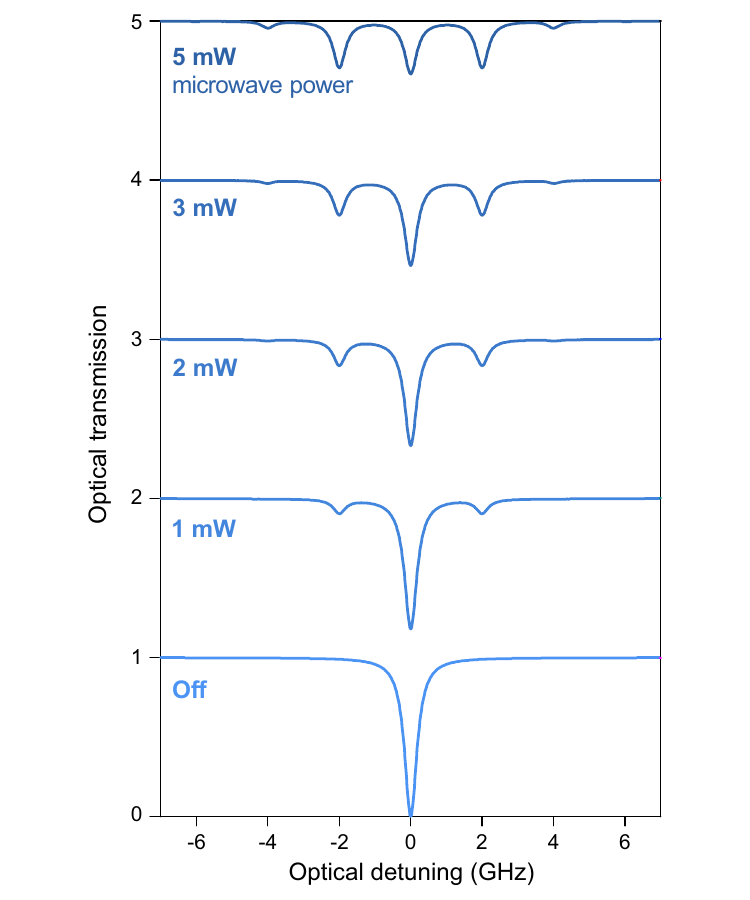}
    \caption{Numerically calculated optical transmission spectra of acousto-optic cavity with microwave input powers from 0 to 5 mW.}
    \label{fig:simulatedTrans}
\end{figure}

The numerically-calculated optical transmission spectra (Fig.~\ref{fig:simulatedTrans}) exhibit sidebands in agreement with the experimental measurements in Fig.~3 in the main text. 

\subsection{$S$ parameter of acousto-optic cavity}
The parameter $S_{21}$ is defined as the normalized microwave power generated by the receiving photodetector, which is generated by beating the pump laser and the optical sideband at the photodetector. 
To derive the power in the optical sideband, we decompose the optical amplitude $a$ into a series of sidebands: 
\begin{equation}
     a = \sum_q A_q e^{-i q \Omega_d t},
\end{equation}
where $A_q$ is the amplitude of optical sideband of order $q$. 
At the weak microwave input power (i.e.~$G \ll \kappa$) limit and in the sideband resolved regime (i.e.~$\Omega_m \gg \kappa$), we only consider the first order of optical sidebands, i.e.~ $q=0,\pm 1$ and, for simplicity, we write the amplitude as, $A_0$, $A_+$, $A_-$. 
The Eq.~\ref{equ:optmotion} is thus decomposed into sidebands, 
\begin{equation}
\begin{split}
    0 & = - \left( i \Delta + \frac{\kappa}{2} \right) A_0 - i G \left(A_++A_-\right) + \sqrt{\kappa_e} A_{in} \\
    -i \Omega_m A_+ & = -\left( i \Delta + \frac{\kappa}{2} \right) A_+ - iG A_0 \\
    i \Omega_m A_- & = -\left( i \Delta + \frac{\kappa}{2} \right) A_- - iG A_0,
\end{split}
\label{equ:optEqu}
\end{equation}
where $A_{in}$ is the input optical amplitude.
The solution of Eq.~\ref{equ:optEqu} is given by
\begin{align}
     A_0 &=\frac{ \sqrt{\kappa_e} A_{in}}{ \left( i \Delta + {\kappa}/{2}
     + G^2  \left(   \frac{1}{i\left( \Delta-\Omega_d \right) + \kappa/2} + \frac{1}{i\left( \Delta+\Omega_d \right) + \kappa/2} \right) 
     \right)} \nonumber \\
     & \simeq \frac{ \sqrt{\kappa_e} A_{in}}{ \left( i \Delta + {\kappa}/{2} \right)} \\
     A_+ &=\frac{-i G A_0}{i\left( \Delta-\Omega_d \right) + \kappa/2} \\
     A_- &=\frac{-i G A_0}{i\left( \Delta+\Omega_d \right) + \kappa/2}
\end{align}

For the scenario the pumping laser is blue detuned from the optical resonance by the acoustic resonant frequency 
($\Delta= - \Omega_m$), and the microwave input is on resonant with the acoustic mode ($\Omega_d = \Omega_m$), the in-cavity optical amplitude for the enhanced sideband $A_-$ is given by,
\begin{equation}
    A_- = \frac{- i G \sqrt{\kappa_e} A_{in}}{\left(-i\Omega_m + \kappa/2\right) \kappa/2},
\end{equation}
where acousto-optic coupling strength $G = g_0 B = 2 g_0 B_{in} \sqrt{\gamma_e} / \gamma$.

Since the pump laser is detuned from the resonant, the transmitted amplitude of the pump laser is close to the input $A_{in}$.
The output microwave voltage $U$ from the photodetector caused by the beating between the transmitted pump laser and the generated optical sideband given by
\begin{equation}
\begin{split}
    U &= R_{PD}\; \hbar \omega \left| \sqrt{\kappa_e}A_- A_{in} \right| \\
      &= R_{PD} \frac{G \kappa_e I_{opt}}{\kappa \sqrt{\Omega_m^2 + \kappa^2/4} /2 } \\
      &\simeq R_{PD} \frac{G \kappa_e I_{opt}}{\Omega_m \kappa/2 },
\end{split}
\end{equation}
where optical power $I_{opt} = \hbar \omega_0  A_{in}^2 $, $\omega_0$ is the optical frequency, and $R_{PD}$ is the response of the photodetector in the unit of $\mathrm{V/W}$. 
The output microwave power is then given by 
\begin{equation}
\begin{split}
    P_{out} &= \frac{U^2}{2 R_{load}}  \\
            &= \frac{ 2 G^2 \kappa_e^2 R_{PD}^2 I_{opt}^2}{\Omega_m^2 \kappa^2 R_{load} },
\end{split}
\end{equation}
where $R_{load} = 50\;\Omega$ is the impedance of the network analyzer. 
The opto-acoustic transmission $S_{21}$ is given by 
\begin{equation}
\begin{split}
    S_{21}&= P_{out} / P_{in} \\
          &= \frac{8 g_0^2 \gamma_e \kappa_e^2 R_{PD}^2 I_{opt}^2 }{\hbar \gamma^2 \Omega_m^3 \kappa^2 R_{load} }
\label{equ:S21}
\end{split}
\end{equation}

\subsection{Estimation of acousto-optic single-photon coupling strength $g_0$ from experimental measurements}

Using the experimental results of the acousto-optic cavity (Fig.~3 in the main text), we can extract the single-photon coupling strength $g_0$ using Eq.~\ref{equ:S21}. 
Taking the insertion loss of the chip into account, the input optical power $I_{opt}$ in Eq.~\ref{equ:S21} is replaced by the power received at the photodetector $I_{rec}$.

We extract the acousto-optic coupling strength $g_0 = 1.1$ kHz from the experimental results shown in Fig.~3 and summarized in Table \ref{tab:g0exp}.
We note this experimentally-extracted $g_0$ is in good agreement with the numerically-simulated value (TE mode, 2.17 GHz) in Table~\ref{tab:VpiLpara}. 
The discrepancy of the acoustic resonant frequency between the numerical simulation and experimental measurement may due to the deviation in LN film thickness and etching depth in fabrication. 

\subsection{Photon number conversion efficiency}
The photon number conversion efficiency $\eta$ relates the number of generated optical sideband photons coupled out of the cavity $\sqrt{\kappa_e}A_-$ to the input microwave photons.
For weak microwave input signals, the conversion efficiency $\eta$ is given by
\begin{equation}
\begin{split}
    \eta &= \left|\frac{\sqrt{\kappa_e}A_-}{B_{in}}\right|^2 \\
    &= \frac{16 g_0^2 \gamma_e  \kappa_e^2 I_\mathrm{opt}}{ \hbar \omega_0 \Omega_m^2 \gamma^2 \kappa^2} \\
    &= \frac{4 g_0^2}{\gamma\,\kappa}\cdot \frac{\kappa_e I_\mathrm{opt}}{\Omega_m^2 \hbar \omega_0}\cdot \frac{2\gamma_e}{\gamma}\cdot \frac{2 \kappa_e}{\kappa} \\
    &= C_0 \cdot n_\mathrm{cav} \cdot \frac{2\gamma_e}{\gamma}\cdot \frac{2 \kappa_e}{\kappa},
\end{split}
\end{equation}
where $C_0 = {4 g_0^2}/(\gamma\kappa)$ is the single-photon cooperativity, $n_\mathrm{cav} = {\kappa_e I_\mathrm{opt}}/(\Omega_m^2 \hbar \omega_0)$ is the intracavity photon number of the blue-detuned pump light, ${2 \kappa_e}/{\kappa}$ is the external coupling efficiency of the optical mode, and  ${2\gamma_e}/{\gamma}$ is the external coupling efficiency of acoustic mode by the IDT.

\begin{table}[ht!]
    \centering
    \caption{Estimation of acousto-optic single-photon coupling strength $g_0$ using experimental results}
    \label{tab:g0exp}
    \begin{tabular}{cc}
        \hline
        Parameter     & Value  \\
        \hline
        Optical mode    & TE \\
        $\omega_0/\, 2\pi$ &  200 THz \\
        $\kappa/\,2\pi$      & 95 MHz      \\
        $ 2 \kappa_e / \kappa$   & 0.3 \\
        $\gamma/\,2\pi$ &      1.28 MHz  \\
        $2 \gamma_e / \gamma$  &   0.34 \\
        $\Omega_m/\,2\pi$  & 2.007 GHz \\
        $R_{PD}$  &  800 V/W \\
        $I_{rec}$  &  0.128 mW \\
        $R_{load}$ & 50 $\Omega$ \\
        $S_{21}$ & -7.5 dB \\
        \hline
        $g_0/\,2\pi$  &  \textbf{1.1 kHz }\\
        \hline
    \end{tabular}
\end{table}

Using the experimentally-extracted values summarized in Table \ref{tab:g0exp}, we estimate a single-photon cooperativity of $C_0 = 4\times10^{-8}$. 
At an optical power of 1 mW, where the intracavity photon number is only about 4,400 due to the large detuning of $\Delta = -\Omega_m$ from the optical resonance, the photon number conversion efficiency is $\eta = 0.0017\%$.

\section{Microwave microscopy of acoustic modes}
\begin{figure}[h]
    \centering
    \includegraphics{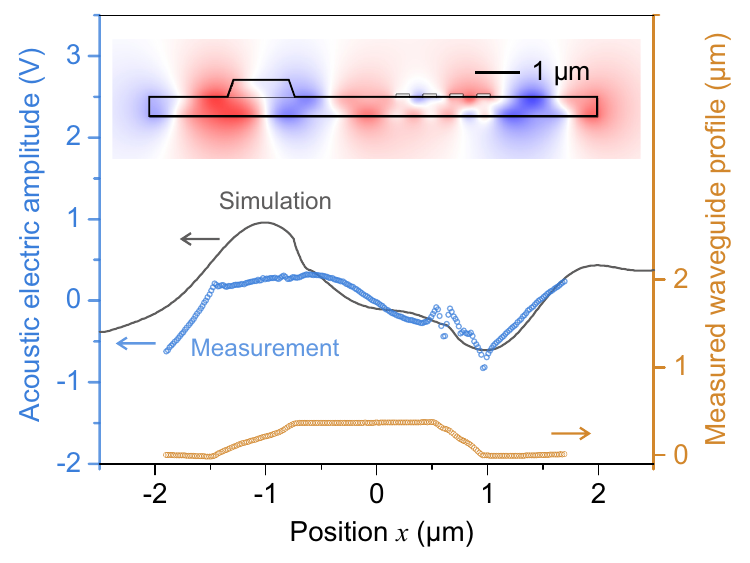}
    \caption{Electric and topographic profiles of an acoustic mode. 
    The electric field amplitude is detected by a scanning probe using transmission-mode microwave impedance microscopy, with a driving signal at 2 GHz on the acoustic resonance.
    Inset: simulated electric field amplitude profile of the acoustic mode.}
    \label{fig:acousticV}
\end{figure}

We experimentally investigate the acoustic mode profiles using transmission-mode microwave impedance microscopy \cite{Zheng2018pnas, Zheng2018prapplied}. 
The working principle is the following -- while the acoustic resonator is driven by a microwave input on the IDT, a probe for atomic force microscopy is scanning over the acoustic resonator and measuring any microwave electric signals. 
The detected signal is mixed with the driving signal to extract the relative amplitude and phase of the acoustic electric field.
The electric amplitude profile of an acoustic mode is obtained on the top surface and in agreement with the numerical simulation (Fig.~\ref{fig:acousticV}). 

\end{document}